\documentclass[
aps,prb,floatfix,
 amsmath,amssymb,
 reprint,%
]{revtex4-2}
\usepackage{chngcntr} 
\usepackage{etoolbox} 
\usepackage{graphicx}
\usepackage{float}
\usepackage[mathlines]{lineno}
\usepackage[font=small]{caption,subcaption}
\usepackage{makecell}
\usepackage{multirow}
\usepackage{orcidlink}
\usepackage{tabularx}
\usepackage{dcolumn}
\usepackage{bm}
\usepackage{tikz}
\makeatletter
\@ifpackageloaded{float}{%
  \AtBeginEnvironment{figure}{\nolinenumbers}%
  \AtBeginEnvironment{figure*}{\nolinenumbers}%
  \AtBeginEnvironment{table}{\nolinenumbers}%
  \AtBeginEnvironment{table*}{\nolinenumbers}%
}{}
\makeatother
\usepackage{mathptmx}
\usepackage{color}
\hypersetup{colorlinks=true,linkcolor=blue}
\usepackage{physics}
\usepackage{ulem}
\newcolumntype{Y}{>{\centering\arraybackslash}X}



\begin{document}

\title{Power-law-anchored residual learning for H-mode energy confinement time in tokamaks: interpolation and parameter-defined extrapolation}

\author{
    Zhaokun Wang\orcidlink{0009-0005-8050-8094}$^{1}$, Tianyuan Liu$^{2,3,4*}$, Jianguo Chen$^{2,3,4}$, Guoyang Shi$^{2,3,4}$, Siqi Ding$^{2,3,4}$, Yuejiang Shi$^{2,3,4}$, Xianmei Zhang$^{1,*}$\\
    \small{$^1$ School of Physics, East China University of Science and Technology, Shanghai 200237, China}\\
    \small{$^2$ Beijing ENN Fusion Energy Science and Technology Co., Ltd., Beijing 101111, China}\\
    \small{$^3$ Beijing Key Laboratory of High Magnetic Field Spherical Torus Fusion Energy, Beijing 101111, China}\\
    \small{$^4$ Hebei Key Laboratory of Compact Fusion, Langfang 065000, China}\\
    \small{$^*$Corresponding authors: liutianyuan@enn.cn; zhangxm@ecust.edu.cn}
}

\begin{abstract}
    Reliable prediction of the energy confinement time $\tau_E$ is essential for magnetic-confinement fusion. Conventional power-law scalings provide constrained extrapolation trends but cannot represent complex nonlinearities, whereas neural networks interpolate accurately but may behave unpredictably outside the training distribution. We propose a unified power-law-anchored residual-learning framework in which a frozen empirical power-law scaling supplies the global trend and a nonlinear model learns only the systematic residual in logarithmic space. PLR-KAN is developed as the primary implementation, while a parameter-matched PLR-MLP serves as a controlled architecture replacement. Using the ITPA DB5.2.3 H-mode confinement database, we evaluate interpolation and parameter-defined held-out cohorts over ten complete training pipelines. PLR-KAN retains near-best interpolation accuracy, achieving $R^2=0.9671\pm0.0027$, while substantially improving the stability of direct KAN under parameter-defined distribution shifts. It outperforms direct KAN across all five non-$\epsilon$ single-parameter-defined cohorts and the core-five joint cohort, reaching $R^2=0.9263\pm0.0157$ in the latter. Results from PLR-MLP further demonstrate that the benefit of power-law anchoring is not specific to KAN, although the effectiveness of residual transfer remains architecture and direction dependent. As an exploratory extension, a Mahalanobis-distance-based prediction-time gate improves stability in selected shifted regions but is not universally beneficial and cannot compensate for missing device or physics-regime coverage. Overall, power-law-anchored residual learning provides a practical balance between nonlinear interpolation capability and empirically constrained extrapolation behavior.

    \quad

\end{abstract}

\maketitle

\section{Introduction}\label{sec:intro}

The energy confinement time, $\tau_E$, is a key parameter characterizing the confinement performance of magnetically confined plasmas~\cite{iter_physics_basis_overview_1999,doyle_chapter_2007}. According to the Lawson criterion~\cite{lawson_criteria_1957,wurzel_progress_2022}, the plasma density, temperature, and energy confinement time jointly determine whether a plasma can approach the condition required for self-sustained burning. For a given heating power, a longer energy confinement time implies a higher plasma stored energy and, consequently, an operating state closer to that required for sustained fusion burn. Therefore, accurate prediction of $\tau_E$ in present devices and future reactors has long been a central issue in magnetic confinement fusion research. It also provides an essential basis for evaluating the fusion gain, auxiliary heating requirements, and accessible operating windows of future devices such as ITER~\cite{doyle_chapter_2007}, CFETR~\cite{wan_overview_2017}, STEP~\cite{chapman_spherical_2024}, and EHL-2~\cite{liang__overview_2025,liu_enns_2024}.

Owing to the extreme complexity of turbulent transport, empirical power-law scaling has long been the main tool for describing and predicting the energy confinement time. This is true both for empirical scaling relations established on individual devices~\cite{ryter_confinement_2001,kaye_energy_2006,valovic_scaling_2009,maslov_energy_2020,kurskiev_energy_2021,jia_scalings_2023} and for multi-device empirical scaling laws based on international databases, such as ITER89-P~\cite{yushmanov_scalings_1990,kaye_iter_1997}, IPB98(y,2)~\cite{confinement_chapter_1999}, and the more recent ITPA20 scaling~\cite{verdoolaege_updated_2021}. These models express $\tau_E$ as a power-law combination of engineering variables or dimensionless normalized parameters. In logarithmic space, they are equivalent to multiple linear regression (MLR). This formulation is physically interpretable because each power-law exponent quantifies the global sensitivity of $\tau_E$ to the corresponding variable, while the prescribed functional form constrains the extrapolated trend and often yields relatively robust predictions. However, a single set of global exponents may also mask regime-dependent confinement physics. For example, NSTX H-mode experiments showed that discharges with boronization plus helium glow discharge conditioning and those with lithium evaporation exhibited different confinement dependences on engineering variables, although the trends could be partly unified by the underlying collisionality dependence~\cite{kaye_dependence_2013}. More recently, high-field spherical-tokamak experiments have provided another example of such regime-dependent behavior. In ST40, a dedicated toroidal-field scan showed a strong confinement dependence, approximately $\tau_E\sim B_T^{2.3}$, for $B_T\leq0.8~\mathrm{T}$, whereas the dependence became very weak or nearly absent at higher magnetic field~\cite{asunta_overview_2026}. A similar saturation of the normalized confinement improvement with increasing $B_T$ was also reported from combined Globus-M2 and ST40 comparisons~\cite{bakharev_recent_2026}. These observations indicate that the magnetic-field dependence in spherical tokamaks may not be represented adequately by a single global power-law exponent, and prescribed power-law forms may be insufficient to capture complex nonlinear or regime-dependent dependencies associated with wall conditioning, collisionality, heating schemes, density regimes, plasma rotation, fast-ion effects, and other physics mechanisms.

To overcome the limitations imposed by a predefined power-law form, data-driven methods such as symbolic regression~\cite{murari_application_2016,murari_robust_2017} and neural networks~\cite{allen_neural_1992} have been introduced for energy confinement time prediction. Symbolic regression searches, within a prescribed operator set, for closed-form scaling expressions without imposing an a priori power-law structure, and thus retains interpretability through explicit analytical formulas. Neural networks, by contrast, provide stronger function approximation capability and often achieve higher interpolation accuracy than conventional power-law models on randomly split datasets. However, improved interpolation accuracy does not necessarily imply reliable extrapolation. Recent analysis of the ITPA H-mode database has similarly shown that increasing model complexity does not necessarily improve extrapolative robustness~\cite{butt_revisiting_2026}. For the design of future devices, the more critical question is whether a model can maintain credible predictions in high-current, high-power, or large-size parameter regimes beyond the training distribution. Previous studies have shown that standard neural networks may produce overconfident or unstable predictions outside the training distribution~\cite{nam_machine_2025}. Therefore, uncertainty quantification~\cite{gao_bayesian_2025}, feature alignment~\cite{nam_machine_2025}, or other extrapolation-constrained strategies are required to improve the reliability of such models.

Cross-device generalization and parameter-defined extrapolation are related but not fully separable in a pooled multi-device database. The held-out cohorts in this work are defined by engineering-parameter thresholds rather than by device labels. However, because device identity and operating ranges are strongly correlated, some cohorts also involve device-domain transfer. These tests should therefore be interpreted as parameter-defined compound shifts rather than pure single-parameter extrapolation. A systematic leave-one-device-out evaluation remains outside the scope of this work.

The Kolmogorov--Arnold network (KAN) is a recently proposed neural-network architecture~\cite{liu_kan_2025}. Unlike the multilayer perceptron (MLP), which employs fixed activation functions at the nodes and learns linear weights, KAN places learnable univariate functions on the edges of the network. This design provides a flexible representation for nonlinear function approximation and offers potential advantages in interpretability. KAN and its variants have been applied to scientific machine learning~\cite{liu_kan_2024}, partial differential equation solving~\cite{patra_physics_2025}, physics-informed learning~\cite{wang_kolmogorov_2025}, and nuclear physics problems~\cite{liu_kolmogorov-arnold_2025}, showing favorable parameter efficiency and interpretability in several benchmark tasks. In the present study, KAN provides a primary high-capacity residual learner with which to test whether strong interpolation can be combined with an empirically constrained extrapolation trend; a parameter-matched MLP provides the controlled architecture replacement.

Motivated by these complementary strengths and weaknesses, this work asks whether a stable empirical trend can be retained while nonlinear learning is confined to the structure that the trend does not explain. Although conventional scaling models are restricted by their prescribed functional form, their linear structure in logarithmic space provides a comparatively constrained trend outside the central data support. Direct neural networks can learn complex in-distribution relationships, but their full prediction is unconstrained in shifted regions. We therefore use the empirical scaling law as a global anchor and restrict the nonlinear model to learning the systematic residual.

This construction defines a unified power-law-anchored residual-learning framework rather than a model tied to one neural architecture. The model first fits and freezes a conventional power-law baseline, then uses a nonlinear function approximator $\mathcal{A}$ to learn the residual correction in logarithmic space. Writing $\hat{y}$ for the predicted $\log\tau_E$, the general form is
\begin{equation}
\hat{y}=f_{\mathrm{PL}}(\mathbf{u})+\Delta_{\mathcal A}(\mathbf{z}),
\label{eq:plr_intro}
\end{equation}
where $f_{\mathrm{PL}}(\mathbf{u})$ supplies the empirical global trend and $\Delta_{\mathcal A}(\mathbf{z})$ represents the learned nonlinear correction. In the primary PLR-KAN implementation, $\mathcal{A}$ is a KAN; in the controlled PLR-MLP replacement, it is a parameter-matched MLP. Both implementations share the power-law baseline, residual target, data partitions, optimization budget, and selection protocol. Their comparison therefore separates the general value of power-law-anchored residual learning from behavior specific to the residual architecture.

Power-law anchoring can substantially reduce an unstable neural-network extrapolation without guaranteeing that the learned residual remains useful in every shifted region. To explore whether this remaining risk can be reduced through a simple prediction-time mechanism, we further examine a Mahalanobis-distance-based attenuation of the already trained and frozen residual. The gated variants, PLR-GATE-KAN and PLR-GATE-MLP, retain the residual most strongly near the training distribution and progressively revert toward the power-law baseline as statistical distance increases. This gate is treated as an exploratory, task-calibrated stability analysis rather than a universal out-of-distribution safeguard. Because neural-network training, data splitting, and hyperparameter selection all carry intrinsic randomness, every interpolation and parameter-defined extrapolation experiment is repeated over ten complete training pipelines, and all results are reported as the mean and sample standard deviation across repetitions.

The main contributions of this work are summarized as follows. (i) We propose a unified power-law-anchored residual-learning framework for H-mode energy-confinement prediction. A frozen empirical power-law scaling provides the global trend, while a nonlinear model learns only the systematic residual in logarithmic space. PLR-KAN is developed as the primary implementation, and a parameter-matched PLR-MLP is constructed as a controlled architecture replacement to distinguish the contribution of the residual framework from that of the KAN architecture. (ii) We systematically evaluate interpolation and parameter-defined extrapolation using the ITPA DB5.2.3 database over ten complete training pipelines. The results show that high interpolation accuracy does not necessarily imply stable or reproducible extrapolation: direct KAN achieves the best interpolation performance but exhibits severe instability in several shifted regions, whereas power-law anchoring substantially improves its performance across all five non-$\epsilon$ single-parameter-defined cohorts and the core-five joint cohort. Comparisons with PLR-MLP further show that the effectiveness of residual correction depends on both the nonlinear architecture and the extrapolation direction. (iii) As an exploratory extension, we apply a Mahalanobis-distance-based prediction-time gate to attenuate the frozen residual in statistically shifted regions. Its beneficial, neutral, and harmful cases demonstrate that simple distance-based residual control can provide additional stability in selected directions, but cannot replace adequate device and physics-regime coverage.

The remainder of this paper follows the same three-part logic. Section~\ref{sec:methods} defines the power-law-anchored residual-learning framework, its KAN and MLP implementations, and the exploratory prediction-time attenuation. Section~\ref{sec:results} evaluates interpolation, parameter-defined extrapolation, architecture--direction dependence, and the beneficial and harmful regimes of gating. Section~\ref{sec:summary} summarizes the three corresponding conclusions, limitations, and future work.

\section{Methods}\label{sec:methods}

This section presents the proposed power-law-anchored residual-learning framework within a unified logarithmic-space formulation. Section~\ref{subsec:plrkan} defines the general residual construction and its primary KAN and controlled MLP implementations. Section~\ref{subsec:gate} then introduces prediction-time residual attenuation as an exploratory extension rather than part of the core framework. Power-law regression, symbolic regression, direct MLP, and direct KAN provide comparison models in Section~\ref{subsec:baselines}, and Section~\ref{subsec:selection} describes the two-stage hyperparameter-selection protocol that excludes held-out-cohort labels from model fitting and selection.

\subsection{Problem formulation and power-law baseline}\label{subsec:setup}

Energy confinement time is modeled at the level of engineering variables. The inputs are the nine engineering parameters commonly used in H-mode scaling studies, listed here in the model input order: plasma current $I_p$, toroidal magnetic field $B_T$, line-averaged electron density $\bar{n}_e$, loss power $P_L$, major radius $R$, elongation $\kappa$, inverse aspect ratio $\epsilon=a/R$ (with $a$ the minor radius), triangularity term $(1+\delta)$ following the standard form used in the ITPA20 scaling~\cite{verdoolaege_updated_2021}, and effective ion mass $M_{\mathrm{eff}}$. The loss power is taken as the net power matched to the thermal confinement time, $P_L=P_{\mathrm{abs}}-\mathrm{d}W/\mathrm{d}t$ (the PLTH column in DB5.2.3), consistent with the regressor used in IPB98(y,2)/ITPA20. The prediction target is the thermal energy confinement time $\tau_E$.

To remain consistent with conventional power-law confinement scaling, the power-law and neural-network formulations use logarithmic variables, and all physical quantities entering the transformation are required to be strictly positive:
\begin{equation}
y=\log \tau_E,\qquad u_i=\log x_i,\qquad i=1,\ldots,9,
\label{eq:log}
\end{equation}
where $x_i$ is the $i$th engineering input and $\log$ denotes the natural logarithm. All neural-network fitting and PLR hyperparameter selection are performed in this logarithmic space through the training loss of Eq.~(\ref{eq:plr_loss}) and the selection score of Eq.~(\ref{eq:score}). PySR candidates are selected separately using validation RMSE in the original $\tau_E$ space. All $R^2$, RMSE, and MAE values reported in Section~\ref{sec:results} are likewise evaluated after transforming predictions back to the original $\tau_E$ space, so that RMSE and MAE (in seconds) are directly interpretable as physical confinement-time errors, while $R^2$ is dimensionless. The conventional power-law scaling $\tau_E = C\prod_i x_i^{\alpha_i}$ reduces in logarithmic space to a linear regression, whose coefficients $\{\beta_i\}$ are obtained by ordinary least squares (OLS):
\begin{equation}
f_{\mathrm{PL}}(\mathbf{u})=\beta_0+\sum_{i=1}^{9}\beta_i u_i .
\label{eq:pl}
\end{equation}
This baseline is simple and provides a comparatively stable extrapolation trend, while belonging to the same family as empirical scalings such as ITPA20. It therefore serves as a reference for attributing the improvement of neural-network models to additional nonlinear representation capability rather than to a change in the assumed functional form.

For the neural-network models, the input is standardized using training-subset statistics,
\begin{equation}
z_i=\frac{u_i-\mu_i^{\mathrm{train}}}{\sigma_i^{\mathrm{train}}},
\label{eq:standard}
\end{equation}
where $\mu_i^{\mathrm{train}}$ and $\sigma_i^{\mathrm{train}}$ are estimated only from the training subset of the current split. To avoid leakage from the validation subset, random test subset, or parameter-defined held-out cohort, all preprocessing parameters (standardization mean and standard deviation, power-law regression coefficients, residual-scaling parameters, and Mahalanobis statistics) are estimated exclusively from the training subset and then applied to the remaining data.

Because neural-network training, random data splitting, and hyperparameter selection all carry intrinsic randomness, every interpolation and parameter-defined extrapolation experiment is repeated over ten complete training pipelines, indexed by $k=0,\ldots,9$. In each pipeline, the non-held-out data are re-split, the networks are re-initialized, the symbolic-regression search is re-run, and all hyperparameters are re-selected; the held-out cohorts themselves are fixed by their physical definitions (Section~\ref{subsec:data}) and do not change across pipelines. If $q_k$ denotes a metric obtained in pipeline $k$, the reported center and dispersion are the arithmetic mean and the sample standard deviation,
\begin{equation}
\bar q=\frac{1}{10}\sum_{k=0}^{9}q_k,
\qquad
s_q^2=\frac{1}{9}\sum_{k=0}^{9}\left(q_k-\bar q\right)^2,
\label{eq:multiseed}
\end{equation}
and all values in Section~\ref{sec:results} are written as $\bar q\pm s_q$. Because the held-out cohorts are fixed, this dispersion quantifies complete-pipeline sensitivity (remaining-data split, initialization, search, and selection) for a given cohort; it does not represent independent device- or population-sampling uncertainty. For gated versus ungated predictions, $\Delta R_k^2=R^2_{\mathrm{gated},k}-R^2_{\mathrm{ungated},k}$ is formed within the same pipeline, separating the paired effect from endpoint scatter. For every paired gate comparison, we report the number of improved pipelines, median paired improvement, and run-to-run stability ratio. For the primary high-$B_T$ PLR-KAN comparison, we additionally report a two-sided percentile-bootstrap 95\% confidence interval for the mean paired improvement, obtained from 100,000 resamples of the ten paired differences with a fixed random seed. The improvement count and run-to-run stability ratio are
\begin{equation}
N_+=\sum_{k=0}^{9}\mathbb{I}\!\left(\Delta R_k^2>0\right),
\qquad
\eta_{\mathrm{SD}}=\frac{s\!\left(R^2_{\mathrm{gated}}\right)}{s\!\left(R^2_{\mathrm{ungated}}\right)}.
\label{eq:gate_stability}
\end{equation}
Thus, $\eta_{\mathrm{SD}}<1$ indicates that gating reduces the run-to-run variability of the complete pipeline.

\subsection{Power-law-anchored residual framework}\label{subsec:plrkan}

The proposed framework is not restricted to a specific residual architecture. It uses the frozen power-law baseline of Eq.~(\ref{eq:pl}) as the global trend and lets a nonlinear function approximator $\mathcal A$ learn only the logarithmic residual. The general prediction is
\begin{equation}
\hat{y}=f_{\mathrm{PL}}(\mathbf{u})+\Delta_{\mathcal A}(\mathbf{z}),
\qquad \hat{\tau}_E=\exp(\hat{y}),
\label{eq:plr_pred}
\end{equation}
where $\mathbf{u}$ is the logarithmic input vector of Eq.~(\ref{eq:log}), $\mathbf{z}$ is its standardized version from Eq.~(\ref{eq:standard}), and $f_{\mathrm{PL}}$ is fitted by OLS on the training subset and then frozen. In PLR-KAN, $\mathcal A$ is a KAN; in the controlled PLR-MLP replacement, it is a parameter-matched MLP. This shared formulation isolates the value of the residual construction from the choice of nonlinear function class.

For numerical stability, the training-set residual $r=y-f_{\mathrm{PL}}(\mathbf{u})$ is used to estimate a mean $\mu_r$ and standard deviation $\sigma_r$ from the training subset only. The residual learner predicts a standardized correction $\widehat{\Delta}_{\mathcal A}(\mathbf{z})$, which is recovered as
\begin{equation}
\Delta_{\mathcal A}(\mathbf{z})=\mu_r+\sigma_r\,\widehat{\Delta}_{\mathcal A}(\mathbf{z}).
\label{eq:resid_recover}
\end{equation}
The same $\mu_r$ and $\sigma_r$ are applied to the validation, test, and held-out cohorts. Because the OLS baseline includes an intercept, the normal equations force the training-set residual mean $\mu_r$ to zero, so Eq.~(\ref{eq:resid_recover}) reduces to $\Delta_{\mathcal A}(\mathbf{z})=\sigma_r\widehat{\Delta}_{\mathcal A}(\mathbf{z})$. The loss combines a logarithmic-space MSE data term with residual-amplitude regularization:
\begin{equation}
\mathcal{L}=\underbrace{\frac{1}{N}\sum_{j=1}^{N}\big(y_j-\hat{y}_j\big)^2}_{\mathcal{L}_{\mathrm{data}}}
+\lambda_r\underbrace{\frac{1}{N}\sum_{j=1}^{N}\Delta_{\mathcal A}(\mathbf{z}_j)^2}_{\mathcal{L}_{\mathrm{res}}} .
\label{eq:plr_loss}
\end{equation}
Here $\mathcal{L}_{\mathrm{res}}$ constrains the recovered correction in the same logarithmic units as the data term, limiting the extent to which the nonlinear branch can perturb the power-law anchor. The coefficient $\lambda_r$ is selected on validation and direction-matched inner-edge subsets for each cohort and pipeline (Section~\ref{subsec:selection}). No distance factor enters training; the gate of Section~\ref{subsec:gate} is applied only after the residual learner is frozen.

KAN is the primary residual implementation. It parametrizes a layered mapping through learnable univariate edge functions. For an $L$-layer network, the $\ell$th layer is defined by a matrix $\boldsymbol{\Phi}^{(\ell)}=\{\varphi^{(\ell)}_{j,i}\}$ and propagates as
\begin{equation}
z^{(\ell+1)}_{j}=\sum_{i}\varphi^{(\ell)}_{j,i}\!\big(z^{(\ell)}_{i}\big),
\label{eq:kan_layer}
\end{equation}
where each edge function has the ``basis $+$ B-spline'' form
\begin{equation}
\varphi(z)=w_b\,b(z)+w_s\sum_{m=1}^{G+k}c_m B_m(z),
\quad b(z)=\frac{z}{1+e^{-z}}.
\label{eq:kan_edge}
\end{equation}
Here $B_m$ are the $k$th-order B-spline basis functions on $G$ grid intervals, $b(z)$ is the SiLU base activation, and $\{w_b,w_s,c_m\}$ are learned parameters. PLR-KAN uses $[9,8,1]$ with $G=3$ and $k=3$, giving about $640$ trainable parameters and matching the architecture and optimization settings of direct KAN.

To test whether the benefit arises from the residual construction rather than from KAN alone, the same formulation is implemented with a parameter-matched MLP residual branch. PLR-MLP shares the OLS baseline, preprocessing, residual target, optimization budget, regularization grid, and selection protocol of PLR-KAN; only the residual function class $\mathcal A$ is changed. It is therefore a controlled architecture replacement and a direct test of the first contribution.

\subsection{Prediction-time residual gating}\label{subsec:gate}

The power-law-anchored residual model constitutes the main proposed framework. As an exploratory extension, we examine whether simple distance-dependent attenuation can reduce excessive residual corrections in strongly shifted regions. The attenuation is applied only at prediction time to an already trained and frozen residual branch, yielding PLR-GATE-KAN and PLR-GATE-MLP; it does not modify the power-law baseline, residual-network weights, or training procedure. It changes Eq.~(\ref{eq:plr_pred}) to
\begin{equation}
\hat{y}=f_{\mathrm{PL}}(\mathbf{u})+g(\mathbf{z})\,\Delta_{\mathcal A}(\mathbf{z}),
\label{eq:gate_pred}
\end{equation}
where $\Delta_{\mathcal A}(\mathbf{z})$ is the recovered logarithmic residual of Eq.~(\ref{eq:resid_recover}) and $g(\mathbf{z})\in(0,1]$ is the attenuation factor. The factor is built on the Mahalanobis distance, which provides a covariance-aware measure of statistical displacement and has also
been used for post-hoc out-of-distribution scoring in neural models~\cite{lee_simple_2018}. In the standardized input space, the mean $\boldsymbol{\mu}_{z}$ and covariance $\boldsymbol{\Sigma}_{z}$ are estimated from the training subset only, the covariance is stabilized by a small trace-scaled ridge, $\boldsymbol{\Sigma}_{\mathrm{reg}}=\boldsymbol{\Sigma}_{z}+10^{-6}\,[\mathrm{tr}(\boldsymbol{\Sigma}_{z})/9]\,\mathbf{I}$, and
\begin{equation}
d_M^2(\mathbf{z})=(\mathbf{z}-\boldsymbol{\mu}_{z})^{\top}\boldsymbol{\Sigma}_{\mathrm{reg}}^{-1}(\mathbf{z}-\boldsymbol{\mu}_{z}),
\quad s_M=\frac{d_M^2}{D},
\label{eq:mahal}
\end{equation}
where $D=9$ is the number of input variables. The gate takes an exponential attenuation form,
\begin{equation}
g(\mathbf{z})=\exp\!\big[-\rho\,s_M(\mathbf{z})\big].
\label{eq:gate}
\end{equation}
When $\rho=0$, $g\equiv1$ and the model reduces to the ungated PLR prediction; for $\rho>0$, the residual is attenuated more strongly as statistical distance increases, and the prediction reverts toward the power-law trend. The strength $\rho$ is selected separately for each predefined held-out cohort and residual branch on a direction-matched inner-edge subset drawn from the training data (Section~\ref{subsec:selection}); scanning or applying $\rho$ involves no backpropagation and changes no trained parameter. Because $\rho$ is calibrated for a known direction using inner-edge labels, this mechanism is task-calibrated rather than a general unsupervised detector of arbitrary distribution shifts. Mahalanobis distance is used instead of a component-wise distance because parameter-defined cohorts in a pooled tokamak database generally involve correlated changes in machine size, magnetic field, plasma current, density, and heating power. The same distance definition is reused in Section~\ref{subsec:data} to construct the core-five joint held-out cohort, there restricted to the five-dimensional core subspace.

To test whether Mahalanobis distance is actually informative of residual reliability, rather than assuming this relation from the gate construction, we define the per-sample excess residual error in logarithmic space as
\begin{equation}
q_i=\left(y_i-\hat y_{\mathrm{PLR},i}\right)^2
-\left(y_i-\hat y_{\mathrm{PL},i}\right)^2,
\label{eq:excess_residual_error}
\end{equation}
where $y_i=\log\tau_{E,i}$. Positive $q_i$ indicates that the learned residual degrades the power-law prediction, whereas negative $q_i$ indicates a beneficial correction. For the representative high-$B_T$ and high-$R$ PLR-KAN cohorts, we compute within each complete training pipeline the Spearman correlation between $q_i$ and $s_M$, divide the held-out samples into five equal-frequency bins according to $s_M$, and calculate the mean $q_i$ and harmful-sample fraction $P(q_i>0)$ in each bin. The correlations and bin summaries are then reported as mean $\pm$ sample standard deviation over the ten pipelines. This post-hoc diagnostic uses only saved predictions and training-derived Mahalanobis statistics; it does not enter model fitting, hyperparameter selection, or gate calibration.

\subsection{Baseline models and training}\label{subsec:baselines}

Power-law regression (PL) is the OLS result of Eq.~(\ref{eq:pl}), used as the empirical power-law reference baseline. Symbolic regression (PySR) uses evolutionary optimization to search for an analytical expression within the operator library $\{+,-,\times,/,\ \mathrm{square},\mathrm{cube},\exp,\log|\cdot|\}$, maintaining a Pareto front of complexity versus loss; candidates on the Pareto front are selected by the validation RMSE in the original $\tau_E$ space (iterations $400$, populations $16$, maxsize $25$). MLP and KAN serve as purely data-driven baselines without any confinement-scaling prior; for a fair comparison their trainable parameter counts are matched to about $640$: the MLP is $[9,20,20,1]$ with SiLU activations, and the KAN is $[9,8,1]$ with the spline edges of Eq.~(\ref{eq:kan_edge}).

As defined in Section~\ref{subsec:plrkan}, PLR-MLP is the controlled architecture replacement for PLR-KAN; the two differ only in the residual function class. All neural networks use the same optimization budget: Adam $1000$ steps $+$ LBFGS $150$ steps, learning rate $10^{-3}$, with MSE loss (Eq.~(\ref{eq:plr_loss}) for the PLR models). In the main experiments of Section~\ref{sec:results}, no L1 or entropy regularization is imposed on KAN ($\lambda_1=\lambda_{\mathrm{entropy}}=0$), so that its full representation capability can be compared fairly with that of MLP; regularization and pruning are used only in Section~\ref{subsec:pruning} for an exploratory inspection of one residual branch. Table~\ref{tab:nn} summarizes the matched neural-network settings.

\begin{table}[h]
\caption{Matched settings of the MLP and KAN families. The same architecture is used either as a direct predictor or as the residual branch of the corresponding PLR model.}
\label{tab:nn}
\begin{ruledtabular}
\begin{tabular}{lll}
Category & MLP family & KAN family \\
\hline
Architecture & $[9,20,20,1]$ & $[9,8,1]$, $G=3$, $k=3$ \\
Nonlinearity & SiLU node & B-spline edges \\
Parameters & $\sim 640$ & $\sim 640$ \\
Direct predictor & MLP & KAN \\
PLR residual branch & PLR-MLP & PLR-KAN \\
\hline
\multicolumn{3}{l}{Optimization: Adam $1000$ $+$ LBFGS $150$, lr $=10^{-3}$; loss: MSE} \\
\end{tabular}
\end{ruledtabular}
\end{table}

\subsection{Hyperparameter selection}\label{subsec:selection}

Because this work reports both random-interpolation and parameter-defined extrapolation results, a unified protocol is applied to all models to avoid task-dependent selection bias and to prevent held-out-cohort information from entering training or selection. The only cases involving genuine free hyperparameter selection are the PLR models (the residual-amplitude coefficient $\lambda_r$) and their gated variants (additionally the gate attenuation strength $\rho$). The protocol is repeated independently in every complete training pipeline, for every predefined held-out cohort, and for each residual branch.

For parameter-defined extrapolation, a randomly sampled validation subset alone cannot identify candidates that fit well in-distribution but deteriorate near the relevant boundary, because the validation subset and held-out cohort do not overlap in that direction. We therefore construct an inner-edge subset $\mathcal{D}_{\mathrm{edge}}$ inside the official training subset $\mathcal{D}_{\mathrm{train}}$, used during selection to mimic samples approaching the boundary without entering the held-out cohort. Its construction matches the definition of that cohort: for single-parameter-defined cohorts it is the top $10\%$ of the training subset in the corresponding physical parameter; for the core-five cohort, the mean vector and covariance matrix of the core subspace are re-estimated using only the training subset, and the top $10\%$ by Mahalanobis score are selected. In the first stage, candidate models are fitted only on $\mathcal{D}_{\mathrm{train}}\setminus\mathcal{D}_{\mathrm{edge}}$ for each $\lambda_r$ in the search grid and evaluated simultaneously on the validation subset and on $\mathcal{D}_{\mathrm{edge}}$. Writing $E_{\mathrm{val}}$ and $E_{\mathrm{edge}}$ for the logarithmic-space RMSE on the two subsets and $E^{\mathrm{PL}}_{\mathrm{edge}}$ for the inner-edge error of the power-law baseline fitted on the same data, the selection score is
\begin{equation}
\mathrm{score}=E_{\mathrm{val}}
+0.5\,E_{\mathrm{edge}}
+\max\!\big(0,\ E_{\mathrm{edge}}-E^{\mathrm{PL}}_{\mathrm{edge}}\big).
\label{eq:score}
\end{equation}
The first term measures in-distribution generalization; the second measures inner-edge boundary error, where the weight $0.5$ is fixed a priori to balance in-distribution accuracy against boundary robustness; the third is a one-sided penalty applied when the candidate model has a larger inner-edge error than the power-law baseline, encoding the prior that a nonlinear correction should not perform worse than the power-law anchor when approaching the predefined direction.

In the second stage, for each held-out cohort, the selection-stage network at the chosen $\lambda_r$ is frozen, and the Mahalanobis statistics of Eq.~(\ref{eq:mahal}) are fitted on the same fitting subset. The gate strength is selected from
$\rho\in\{0,0.01,0.02,0.05,0.1,0.2,0.5,1,2\}$ by minimizing $E_{\mathrm{edge}}(\rho)$ on the direction-matched inner-edge subset, with ties resolved in favor of the smaller
candidate. This stage uses inner-edge labels only; it is therefore a supervised boundary calibration for a known extrapolation direction rather than an unsupervised detector of arbitrary shifts.

After $\lambda_r$ and $\rho$ are fixed, the selection-stage network is discarded, the final ungated PLR model is retrained on the complete training subset with the selected $\lambda_r$, and all preprocessing and Mahalanobis statistics are re-estimated on that subset. The selected $\rho$ is applied only when the final gated predictions are generated. The held-out cohort is used only after the final model is fixed, solely for reporting.

The residual-amplitude search range is $\lambda_r\in\{0,10^{-3},10^{-2},0.03,0.1,0.3,1,3\}$
for both the random-interpolation and parameter-defined extrapolation experiments. For random interpolation, there is no predefined extrapolation direction, so a generic top-$10\%$ nine-feature training-edge subset is used in place of the direction-matched edge, and no gate is applied. Here $\lambda_r=0$ removes the residual-amplitude penalty, whereas $\rho=0$ recovers the ungated PLR prediction. Both grids deliberately extend beyond the typically selected values, and the largest candidates ($\lambda_r=3$ and
$\rho=2$) are never selected, so the reported selections are not truncated by the grid boundary. Here $\lambda_r$ refers only to the residual-amplitude regularization and is independent of the internal KAN L1/entropy coefficients; the latter are kept at zero in the main experiments and activated only in Section~\ref{subsec:pruning} for the
pruning analysis.

\section{Results and discussion}\label{sec:results}

\subsection{Dataset and evaluation design}\label{subsec:data}

Model training, validation, and extrapolation tests use the STD5 standard subset of the ITPA DB5.2.3 H-mode energy confinement database~\cite{verdoolaege_updated_2021}. This database accumulates long-term experimental data from international tokamaks, covering diverse operating regimes, machine sizes, parameter ranges, plasma shapes, heating schemes, and wall-material conditions. The STD5 subset consists of high-quality H-mode discharges suitable for scaling analysis, including both ELMy and ELM-free samples. After removing entries with missing variables, the final dataset contains $N=7546$ samples from $18$ devices. Table~\ref{tab:device} summarizes the device composition: JET (including the ITER-like wall) and ASDEX Upgrade (AUG, including tungsten-wall operation) account for large fractions, indicating that the bias introduced by the non-uniform device distribution should be considered in subsequent training and extrapolation assessment.

\begin{table}[h]
\caption{Device composition of the final processed dataset ($N=7546$, $18$ devices).}
\label{tab:device}
\begin{ruledtabular}
\begin{tabular}{lccccc}
Device & $N$ & Frac. & Device & $N$ & Frac. \\
\hline
JET & 3101 & 41.1\% & CMOD & 82 & 1.1\% \\
AUG & 2141 & 28.4\% & MAST & 39 & 0.5\% \\
ASDEX & 575 & 7.6\% & TUMAN3M & 36 & 0.5\% \\
DIII-D & 502 & 6.7\% & COMPASS & 17 & 0.2\% \\
JFT2M & 349 & 4.6\% & TCV & 17 & 0.2\% \\
NSTX & 230 & 3.0\% & TDEV & 10 & 0.1\% \\
PBXM & 214 & 2.8\% & START & 8 & 0.1\% \\
PDX & 119 & 1.6\% & T10 & 4 & $<0.1\%$ \\
JT60U & 100 & 1.3\% & TFTR & 2 & $<0.1\%$ \\
\end{tabular}
\end{ruledtabular}
\end{table}

To distinguish in-distribution interpolation capability from robustness under predefined distribution shifts, two complementary splitting strategies are used. \textit{(1) Stratified random split (interpolation).} Because $\tau_E$ spans roughly $1~\mathrm{ms}$ to $1~\mathrm{s}$, a simple random split can make the target distributions of the subsets inconsistent. Using $y=\log\tau_E$ as the stratification variable, all samples are sorted into $10$ equal-frequency bins and, within each bin, assigned to $\mathrm{train}\!:\!\mathrm{validation}\!:\!\mathrm{test}=70\%\!:\!10\%\!:\!20\%$. This produces similar $\log\tau_E$ distributions across subsets, and the stratified split is regenerated in every complete training pipeline. \textit{(2) Parameter-defined held-out cohorts (extrapolation).} Each held-out cohort is defined first and remains fixed across pipelines; the remaining samples are randomly divided into train/validation/test at $7\!:\!1\!:\!1$ in each pipeline. For the relatively continuous $I_p,\bar{n}_e,P_L$, the cohort is the top $10\%$ of the corresponding variable over the full database, i.e. $\mathcal{D}_{\mathrm{held}}^{(x_i)}=\{j:x_{i,j}\ge q_{90}(x_i)\}$. For $B_T,R,\epsilon$, which show clear device clustering or discrete-boundary behavior, fixed physical thresholds are used: $B_T>4.0~\mathrm{T}$, $R>3.0~\mathrm{m}$, and $\epsilon>0.5$, corresponding to high-field, large-size, and high-inverse-aspect-ratio (spherical-tokamak-like) regimes. The core-five cohort instead uses the engineering variables $\{I_p,B_T,P_L,R,\bar{n}_e\}$, computes $S_5(\mathbf{z})=\tfrac15 d_M^2(\mathbf{z}_{\mathrm{core5}})$ from Eq.~(\ref{eq:mahal}) in this five-dimensional subspace, and holds out the top $10\%$. The $S_5$ defining this cohort is computed from full-database, input-only statistics before splitting and therefore introduces no target-information leakage; it is deliberately distinct from the inner-edge $S_5$ of Section~\ref{subsec:selection}, which uses training-subset statistics only.

\begin{figure}[h]
    \centering
    \includegraphics[width=.9\linewidth]{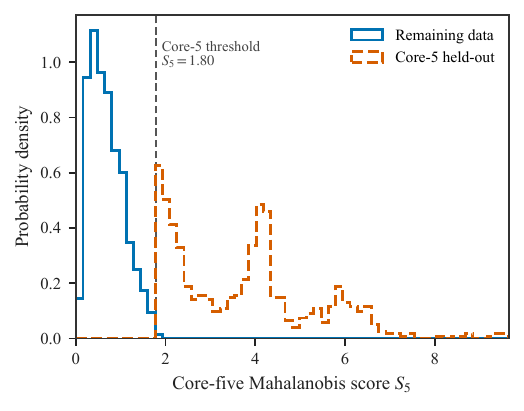}
    \caption{Probability-density histograms (each normalized to unit area) of the core-five Mahalanobis score $S_5=\tfrac15 d_M^2(\mathbf{z}_{\mathrm{core5}})$, computed on the five core engineering variables $\{I_p,B_T,P_L,\bar{n}_e,R\}$ using full-database statistics. The remaining data (train$+$validation$+$test, solid) concentrate at low $S_5$, whereas the core-five held-out subset (top $10\%$ by $S_5$, dashed) lies beyond the split threshold and is clearly separated. Line styles distinguish the subsets so the figure remains readable in grayscale.}
    \label{fig:score}
\end{figure}

\begin{figure*}[t]
    \centering
    \includegraphics[width=.95\textwidth]{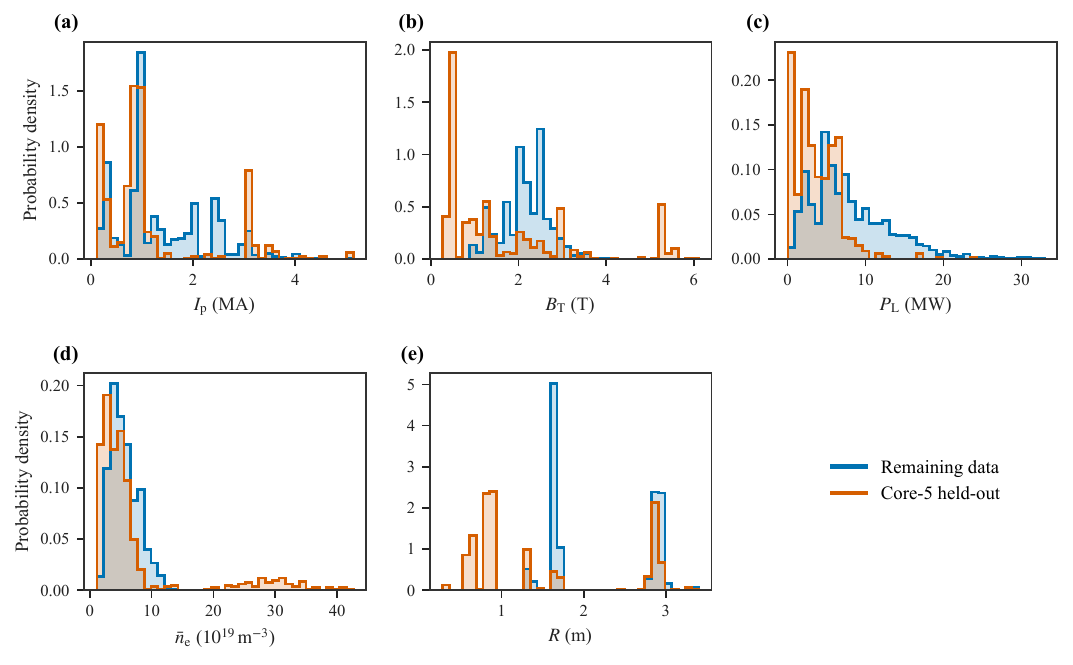}
    \caption{Marginal distributions of the five core engineering parameters [(a) $I_p$, (b) $B_T$, (c) $P_L$, (d) $\bar{n}_e$, (e) $R$] for the core-five held-out cohort. Each histogram is normalized to unit area. Orange curves correspond to the held-out cohort; blue curves represent the remaining in-distribution data (train$+$validation$+$test). The cohort is not uniformly high in every variable but reflects a correlated multi-parameter shift.}
    \label{fig:margin}
\end{figure*}

The single-parameter-defined and core-five cohorts are structurally complementary: the former probe model behavior when one key engineering parameter enters a high-value or physical-boundary region, whereas the latter probes generalization under a joint shift of the core parameters. Figures~\ref{fig:score} and~\ref{fig:margin} show that the remaining in-distribution data concentrate at low $S_5$, whereas the core-five cohort lies at higher $S_5$ and is clearly separated. The cohort is not uniformly high in all five variables; for example, part of it combines high $B_T$ with low $R$ and low $\bar{n}_e$, while $I_p$ and $P_L$ are not uniformly high. Table~\ref{tab:extrap} lists the sample size and device composition of every cohort. Because engineering parameters and device identity are strongly correlated, these are parameter-defined compound shifts rather than controlled one-variable scans. In particular, the high-$B_T$, high-$R$, and $\epsilon>0.5$ cohorts contain all C-Mod, JT-60U, and spherical-tokamak samples, respectively. Thus, ``single-parameter-defined'' identifies how a cohort is constructed; it does not imply that all remaining variables or device identities stay in-distribution.

\begin{table*}[t]
\caption{Definition, sample size, and device composition of the parameter-defined held-out cohorts.}
\label{tab:extrap}
\begin{ruledtabular}
\begin{tabular}{llcl}
Cohort & Definition & $N_{\mathrm{held}}$ & Device composition \\
\hline
High $I_p$ & top $10\%$ in $I_p$ & 755 & JET (755) \\
High $\bar{n}_e$ & top $10\%$ in $\bar{n}_e$ & 755 & AUG (494), JET (116), CMOD (82), DIII-D (53), others (10) \\
High $P_L$ & top $10\%$ in $P_L$ & 755 & JET (724), AUG (19), JT60U (8), DIII-D (4) \\
High $B_T$ & $B_T>4.0$ T & 86 & CMOD (82), JT60U (2), TFTR (2) \\
High $R$ & $R>3.0$ m & 120 & JT60U (100), JET (20) \\
High $\epsilon$ & $\epsilon>0.5$ & 277 & NSTX (230), MAST (39), START (8) \\
Core-five & top $10\%$ in $S_5$ & 755 & NSTX (230), JET (196), CMOD (82), JFT2M (61), others (186) \\
\end{tabular}
\end{ruledtabular}
\end{table*}

\subsection{Interpolation results}\label{subsec:interp}

\begin{figure*}[t]
    \centering
    \includegraphics[width=.95\textwidth]{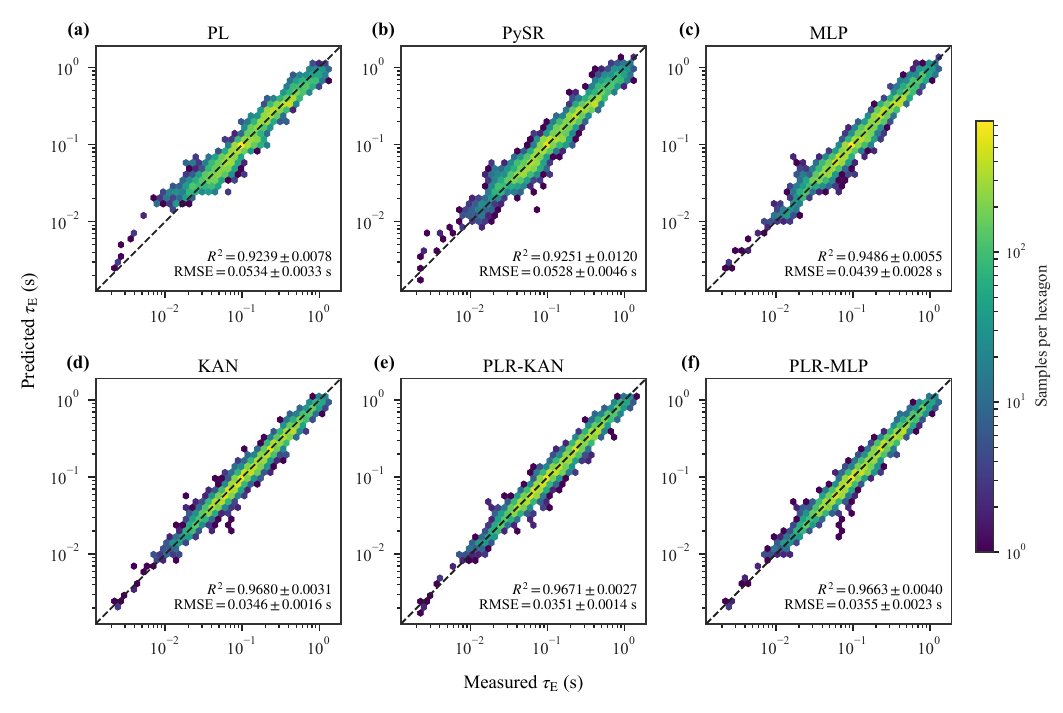}
    \caption{Predicted-versus-experimental thermal energy confinement time $\tau_E$ on the stratified-random test subsets, for (a) PL, (b) PySR, (c) MLP, (d) KAN, (e) PLR-KAN, and (f) PLR-MLP. Test predictions from all ten complete training pipelines are pooled only to visualize density. Both axes are logarithmic; the dashed diagonal marks perfect prediction, and color encodes local sample density. In-plot annotations give $R^2$ (dimensionless) and RMSE (seconds) as mean $\pm$ sample standard deviation; the values match Table~\ref{tab:interp}.}
    \label{fig:interp_scatter}
\end{figure*}

We first compare interpolation performance on the stratified-random test subsets. Figure~\ref{fig:interp_scatter} shows predicted-versus-experimental scatter plots, and Table~\ref{tab:interp} summarizes $R^2$, RMSE, and MAE in the original $\tau_E$ space over ten complete training pipelines. The gated variants are not listed: $\rho$ requires a predefined direction for inner-edge calibration, so the gate does not define a separate model for the in-distribution interpolation test and is evaluated only in Sections~\ref{subsec:extrap}--\ref{subsec:gate_result}.

Fitting Eq.~(\ref{eq:pl}) by OLS on each stratified-random training subset yields an empirical power-law scaling; averaging the coefficients over the ten repetitions gives
\begin{equation}
\begin{aligned}
\tau_E={}&0.0798\,I_p^{1.059}B_T^{0.120}
\bar{n}_e^{0.113}P_L^{-0.654}\\
&{}\times R^{1.450}\kappa^{0.194}\epsilon^{0.033}
(1+\delta)^{0.194}M_{\mathrm{eff}}^{0.302}.
\end{aligned}
\label{eq:pl_fit}
\end{equation}
The power-law baseline achieves $R^2=0.9239\pm0.0078$ and RMSE $=0.0534\pm0.0033$, showing that a global empirical scaling captures most of the variation in H-mode $\tau_E$ but leaves systematic nonlinear structure. PySR performs similarly ($R^2=0.9251\pm0.0120$, RMSE $=0.0528\pm0.0046$). Direct neural predictors are more accurate: MLP reaches $R^2=0.9486\pm0.0055$ and RMSE $=0.0439\pm0.0028$, while KAN gives the best interpolation result, $R^2=0.9680\pm0.0031$ and RMSE $=0.0346\pm0.0016$. The scatter plots in Fig.~\ref{fig:interp_scatter} show the same ordering, with the largest systematic deviations for PL and PySR in the high-$\tau_E$ region and the closest concentration around the diagonal for KAN. Crucially, PLR-KAN ($R^2=0.9671\pm0.0027$, RMSE $=0.0351\pm0.0014$) and PLR-MLP ($R^2=0.9663\pm0.0040$, RMSE $=0.0355\pm0.0023$) nearly match direct KAN. Thus, direct KAN provides the strongest interpolation capability, whereas both PLR implementations retain nearly the same accuracy. This establishes that constraining nonlinear learning to the residual space does not materially sacrifice in-distribution representation capability. It does not, however, establish robustness under parameter-defined shifts, which is examined next.

\begin{table}[h]
\caption{Performance on the stratified-random test subsets over ten complete training pipelines (mean $\pm$ sample standard deviation; original $\tau_E$ space; RMSE and MAE in seconds, $R^2$ dimensionless).}
\label{tab:interp}
\begin{ruledtabular}
\begin{tabular}{lccc}
Model & $R^2$ & RMSE & MAE \\
\hline
Power law & $0.9239\pm0.0078$ & $0.0534\pm0.0033$ & $0.0310\pm0.0012$ \\
PySR      & $0.9251\pm0.0120$ & $0.0528\pm0.0046$ & $0.0307\pm0.0016$ \\
MLP       & $0.9486\pm0.0055$ & $0.0439\pm0.0028$ & $0.0254\pm0.0009$ \\
KAN       & $0.9680\pm0.0031$ & $0.0346\pm0.0016$ & $0.0197\pm0.0007$ \\
PLR-KAN   & $0.9671\pm0.0027$ & $0.0351\pm0.0014$ & $0.0202\pm0.0005$ \\
PLR-MLP   & $0.9663\pm0.0040$ & $0.0355\pm0.0023$ & $0.0207\pm0.0009$ \\
\end{tabular}
\end{ruledtabular}
\end{table}

\subsection{Parameter-defined extrapolation}\label{subsec:extrap}

\begin{figure*}[t]
    \centering
    \includegraphics[width=.95\textwidth]{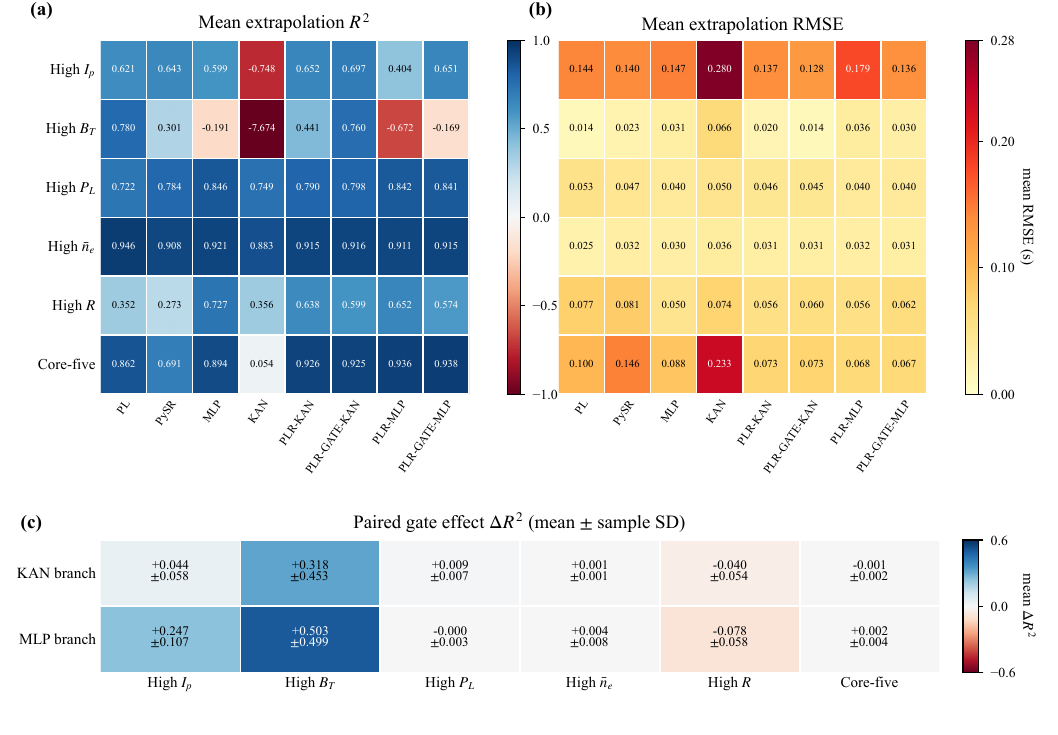}
    \caption{Performance over ten complete training pipelines for five single-parameter-defined held-out cohorts ($I_p,B_T,P_L,\bar{n}_e,R$) and the core-five cohort; (a) Mean $R^2$; (b) Mean RMSE in seconds; (c) Paired prediction-time gate effect $\Delta R^2$ between gated and ungated PLR models. Values are averaged over ten pipelines; error bars denote sample standard deviation.}
    \label{fig:extrap}
\end{figure*}

We first isolate the central contribution of power-law anchoring by comparing the ungated models---PL, PySR, MLP, KAN, PLR-KAN, and PLR-MLP---on six single-parameter-defined cohorts ($I_p$, $B_T$, $P_L$, $\bar{n}_e$, $R$, and $\epsilon$) and the core-five cohort. Figure~\ref{fig:extrap} shows mean $R^2$ and RMSE over ten complete training pipelines for the five non-$\epsilon$ cohorts and core-five, and Table~\ref{tab:extrap_r2} gives the corresponding $R^2$ distributions. Gated predictions and paired gate effects are discussed only in Section~\ref{subsec:gate_result}. The high-$\epsilon$ cohort is treated separately because all eight prediction variants have negative mean $R^2$, indicating a compound device- and physics-regime shift rather than a continuous extension along one parameter. Performance is strongly direction dependent. Most models remain useful for high $\bar{n}_e$ and high $P_L$, whereas high $B_T$ and high $R$ more clearly expose the contrast between the reproducible power-law trend and the variability of flexible predictors.

\begin{table*}[t]
\caption{Mean $R^2$ over ten complete training pipelines for parameter-defined held-out cohorts. Values are reported as mean $\pm$ sample standard deviation. Gated columns share trained weights with the corresponding ungated PLR models. High-$\epsilon$ results are discussed separately because all models yield negative mean $R^2$.}
\label{tab:extrap_r2}
\centering
\scriptsize
\setlength{\tabcolsep}{3.0pt}
\begin{ruledtabular}
\begin{tabular}{lcccccccc}
Cohort & PL & PySR & MLP & KAN & PLR-KAN & \makecell{PLR-GATE\\KAN} & PLR-MLP & \makecell{PLR-GATE\\MLP} \\
\hline
High $I_p$ &
\makecell{$0.6210$\\$\pm0.0050$} &
\makecell{$0.6430$\\$\pm0.0227$} &
\makecell{$0.5989$\\$\pm0.0969$} &
\makecell{$-0.7477$\\$\pm1.6978$} &
\makecell{$0.6524$\\$\pm0.0712$} &
\makecell{$0.6969$\\$\pm0.0426$} &
\makecell{$0.4039$\\$\pm0.1417$} &
\makecell{$0.6510$\\$\pm0.1297$} \\
High $B_T$ &
\makecell{$0.7796$\\$\pm0.0084$} &
\makecell{$0.3012$\\$\pm0.4882$} &
\makecell{$-0.1911$\\$\pm0.5414$} &
\makecell{$-7.6744$\\$\pm16.7215$} &
\makecell{$0.4411$\\$\pm0.5029$} &
\makecell{$0.7595$\\$\pm0.0915$} &
\makecell{$-0.6716$\\$\pm1.0308$} &
\makecell{$-0.1689$\\$\pm0.8106$} \\
High $P_L$ &
\makecell{$0.7216$\\$\pm0.0049$} &
\makecell{$0.7838$\\$\pm0.0477$} &
\makecell{$0.8456$\\$\pm0.0132$} &
\makecell{$0.7489$\\$\pm0.0887$} &
\makecell{$0.7899$\\$\pm0.0597$} &
\makecell{$0.7985$\\$\pm0.0556$} &
\makecell{$0.8416$\\$\pm0.0260$} &
\makecell{$0.8413$\\$\pm0.0249$} \\
High $\bar{n}_e$ &
\makecell{$0.9459$\\$\pm0.0016$} &
\makecell{$0.9083$\\$\pm0.0249$} &
\makecell{$0.9212$\\$\pm0.0122$} &
\makecell{$0.8825$\\$\pm0.0536$} &
\makecell{$0.9152$\\$\pm0.0196$} &
\makecell{$0.9162$\\$\pm0.0199$} &
\makecell{$0.9109$\\$\pm0.0168$} &
\makecell{$0.9147$\\$\pm0.0136$} \\
High $R$ &
\makecell{$0.3523$\\$\pm0.0084$} &
\makecell{$0.2727$\\$\pm0.2524$} &
\makecell{$0.7275$\\$\pm0.0512$} &
\makecell{$0.3557$\\$\pm0.4082$} &
\makecell{$0.6384$\\$\pm0.1844$} &
\makecell{$0.5986$\\$\pm0.1680$} &
\makecell{$0.6518$\\$\pm0.0903$} &
\makecell{$0.5735$\\$\pm0.0716$} \\
Core-five &
\makecell{$0.8618$\\$\pm0.0032$} &
\makecell{$0.6906$\\$\pm0.1569$} &
\makecell{$0.8939$\\$\pm0.0101$} &
\makecell{$0.0536$\\$\pm0.9298$} &
\makecell{$0.9263$\\$\pm0.0157$} &
\makecell{$0.9249$\\$\pm0.0146$} &
\makecell{$0.9355$\\$\pm0.0106$} &
\makecell{$0.9379$\\$\pm0.0081$} \\
\end{tabular}
\end{ruledtabular}
\end{table*}

Direct KAN illustrates why interpolation accuracy alone is insufficient. In the high-$B_T$ cohort, the power law is accurate and reproducible ($R^2=0.7796\pm0.0084$), whereas direct KAN fails severely ($-7.6744\pm16.7215$, with the worst repetition below $-54$); MLP and PySR are also unstable. In the high-$R$ cohort, the power law is reproducible but biased ($0.3523\pm0.0084$), direct MLP improves the mean score to $0.7275\pm0.0512$, and direct KAN again shows a much larger spread ($0.3557\pm0.4082$). The same architecture that gives the best interpolation result can therefore be among the least stable under selected parameter-defined shifts.

Power-law anchoring changes that comparison systematically. PLR-KAN exceeds direct KAN in mean $R^2$ on all five non-$\epsilon$ single-parameter-defined cohorts and on core-five. The clearest gains over the power-law baseline occur for high $R$ and core-five: PLR-KAN raises $R^2$ from $0.3523$ to $0.6384\pm0.1844$ for high $R$, and reaches $0.9263\pm0.0157$ on core-five, compared with $0.8618\pm0.0032$ for PL, $0.8939\pm0.0101$ for MLP, and $0.0536\pm0.9298$ for direct KAN. PLR-MLP reaches $0.9355\pm0.0106$ on core-five, showing that the framework is not specific to KAN. At the same time, PLR-MLP does not uniformly improve upon direct MLP, and the better residual architecture changes by direction. The evidence therefore separates the value of the power-law anchor from the choice of nonlinear residual learner: anchoring is especially valuable for stabilizing KAN, but its quantitative benefit remains architecture and direction dependent.

High $B_T$ defines the remaining limitation of ungated PLR. Anchoring converts direct KAN's severe failure to a positive PLR-KAN mean, $R^2=0.4411\pm0.5029$, but this remains below the accurate power-law baseline and retains substantial run-to-run variability. The additive formulation in Eq.~(\ref{eq:plr_pred}) stabilizes the global trend but does not guarantee that the learned residual stays useful outside its training support. By contrast, high $I_p$ is not a failed-transfer case: PLR-KAN reaches $0.6524\pm0.0712$, above both PL ($0.6210\pm0.0050$) and PySR ($0.6430\pm0.0227$). These two directions show why extrapolation behavior must be evaluated rather than inferred from model capacity or from interpolation ranking.

The $\epsilon>0.5$ cohort is a stronger boundary case. All eight prediction variants have negative mean $R^2$; even PL reaches only $-0.3350\pm0.0630$. Neither residual learning nor Mahalanobis attenuation repairs the failure: PLR-KAN and PLR-GATE-KAN give $-1.3973\pm0.8849$ and $-1.3412\pm0.8823$, while PLR-MLP and PLR-GATE-MLP give $-2.4070\pm1.2626$ and $-1.8853\pm1.0176$. This cohort contains the spherical tokamaks START, MAST, and NSTX and therefore represents a compound device- and physics-regime transfer rather than a continuous extension in inverse aspect ratio alone. Published spherical-tokamak scalings also report stronger toroidal-field and weaker plasma-current dependence than conventional scalings~\cite{kaye_thermal_2021,kaye_energy_2006,valovic_scaling_2009}. The collective failure is thus consistent with missing physical coverage in the training data: neither an unsupported residual nor reversion toward an unsupported baseline can supply the absent regime.

Overall, the second contribution is an empirical result rather than a universal guarantee: high interpolation accuracy does not imply stable parameter-defined extrapolation, and power-law anchoring substantially reduces the most severe instability of direct KAN. PLR-KAN improves upon direct KAN across all five non-$\epsilon$ single-parameter-defined cohorts and core-five, while PLR-MLP confirms that the formulation extends beyond KAN. The remaining differences across architectures and directions, together with the collective high-$\epsilon$ failure, define the boundary of that benefit. The unresolved high-$B_T$ variability motivates the exploratory prediction-time attenuation analyzed next.

\subsection{Residual gating: benefits and limitations}\label{subsec:gate_result}

Motivated by the remaining high-$B_T$ variability, we evaluate the prediction-time gate of Section~\ref{subsec:gate} as an exploratory extension to the PLR framework. The Mahalanobis attenuation in Eq.~(\ref{eq:gate}) changes only how strongly a frozen residual branch modifies the power-law anchor; it does not retrain the residual model. Because $\rho$ is calibrated on a direction-matched inner edge, these results test task-calibrated attenuation for predefined held-out cohorts, not a universal detector of arbitrary distribution shift. Figure~\ref{fig:extrap}(c) shows paired within-pipeline effects, and Table~\ref{tab:gate} reports selected hyperparameters, attenuation strength, paired improvement, and run-to-run stability.

\begin{table*}[!t]
    \caption{Paired effects and stability of the exploratory prediction-time gate over ten complete training pipelines, for each held-out cohort and residual branch. The hyperparameter columns give the modal selected $\lambda_r$ and $\rho$ (tied modes are both listed). $\langle g\rangle_{\mathrm{held}}$ is the mean $\pm$ sample standard deviation, across pipelines, of each pipeline's mean gate value on the fixed held-out cohort. $\Delta R^2$ is the paired within-pipeline effect, $N_+$ counts strict improvements, and $\eta_{\mathrm{SD}}$ is defined in Eq.~(\ref{eq:gate_stability}). A small $\langle g\rangle_{\mathrm{held}}$ means stronger reversion toward the power-law anchor.}
    \label{tab:gate}
    \centering
    \scriptsize
    \setlength{\tabcolsep}{2.5pt}
    \begin{ruledtabular}
    \begin{tabular}{llccccccc}
        Cohort & Branch & modal $\lambda_r$ & modal $\rho$ & $\langle g\rangle_{\mathrm{held}}$ & paired $\Delta R^2$ & median $\Delta R^2$ & $N_+/10$ & $\eta_{\mathrm{SD}}$ \\
        \hline
        \multirow{2}{*}{High $I_p$} & KAN & $0.10$ & $0.50$ & $0.656\pm0.169$ & $+0.0445\pm0.0576$ & $+0.0290$ & $9/10$ & $0.598$ \\
         & MLP & $0.10$ & $0.50$ & $0.692\pm0.172$ & $+0.2471\pm0.1073$ & $+0.2266$ & $10/10$ & $0.915$ \\
        \multirow{2}{*}{High $B_T$} & KAN & $0.30$ & $0.50$ & $0.374\pm0.328$ & $+0.3185\pm0.4531$ & $+0.0919$ & $9/10$ & $0.182$ \\
         & MLP & $0.10$ & $0.10$ & $0.666\pm0.231$ & $+0.5027\pm0.4986$ & $+0.4343$ & $8/10$ & $0.786$ \\
        \multirow{2}{*}{High $P_L$} & KAN & $0.10$ & $0.10$ & $0.911\pm0.045$ & $+0.0086\pm0.0067$ & $+0.0073$ & $9/10$ & $0.930$ \\
         & MLP & $0.10$ & $0.10$ & $0.897\pm0.035$ & $-0.0003\pm0.0030$ & $-0.0005$ & $4/10$ & $0.961$ \\
        \multirow{2}{*}{High $\bar{n}_e$} & KAN & $0,\ 0.03$ & $0$ & $0.980\pm0.023$ & $+0.0010\pm0.0011$ & $+0.0007$ & $6/10$ & $1.012$ \\
         & MLP & $0,\ 0.03$ & $0$ & $0.980\pm0.030$ & $+0.0038\pm0.0079$ & $0.0000$ & $4/10$ & $0.809$ \\
        \multirow{2}{*}{High $R$} & KAN & $0.30$ & $0.50$ & $0.566\pm0.210$ & $-0.0398\pm0.0539$ & $-0.0383$ & $2/10$ & $0.911$ \\
         & MLP & $0.30$ & $0.50$ & $0.460\pm0.172$ & $-0.0783\pm0.0575$ & $-0.0765$ & $0/10$ & $0.793$ \\
        \multirow{2}{*}{Core-five} & KAN & $0.10$ & $0.05$ & $0.735\pm0.169$ & $-0.0014\pm0.0023$ & $-0.0006$ & $2/10$ & $0.933$ \\
         & MLP & $0.30$ & $0.05,\ 0.10$ & $0.734\pm0.187$ & $+0.0024\pm0.0036$ & $+0.0013$ & $8/10$ & $0.765$ \\
    \end{tabular}
    \end{ruledtabular}
\end{table*}

\begin{figure*}[!t]
    \centering
    \includegraphics[width=.92\textwidth]{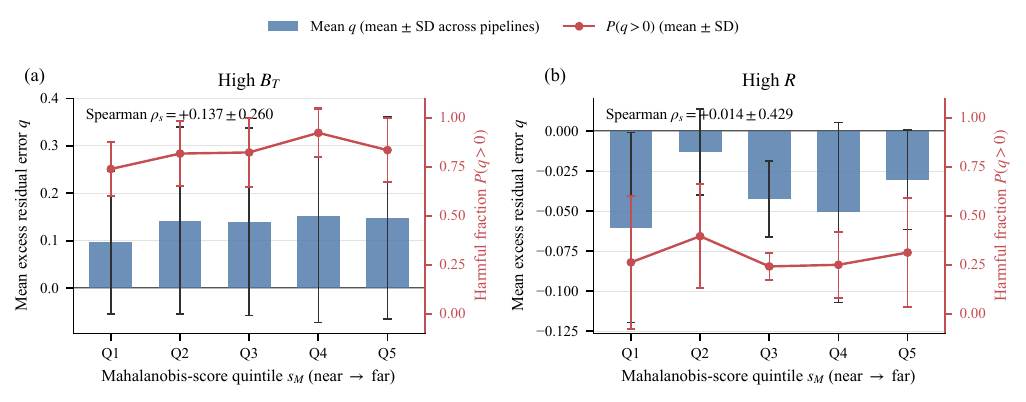}
    \caption{Distance-resolved residual-risk diagnostic for ungated PLR-KAN in two representative held-out cohorts. The per-sample excess residual error $q_i$ is defined by Eq.~(\ref{eq:excess_residual_error}), so $q_i>0$ means that the learned residual correction is harmful relative to the power-law prediction. Within each of the ten complete training pipelines, samples are divided into five equal-frequency bins by the training-derived Mahalanobis score $s_M$. Bars show mean $q_i$, and lines show the harmful-sample fraction $P(q_i>0)$; centers and error bars summarize pipelines. High $B_T$ shows a weak and nonmonotonic positive association, whereas high $R$ shows no consistent distance--error association and retains a beneficial mean residual in every bin. This post-hoc diagnostic uses saved predictions only and does not enter fitting or selection.}
    \label{fig:distance_reliability}
\end{figure*}

High $B_T$ is the strongest positive case. Ungated PLR-KAN reaches $R^2=0.4411\pm0.5029$, and the selected gate raises it to $0.7595\pm0.0915$ with $\langle g\rangle_{\mathrm{held}}=0.374\pm0.328$. The paired improvement is $\Delta R^2=+0.3185\pm0.4531$, with median $+0.0919$ and a percentile-bootstrap 95\% confidence interval of $[0.0827,\,0.6104]$ for the mean; 9 of 10 paired pipelines improve. The run-to-run standard deviation decreases by approximately 82\%, from $0.5029$ to $0.0915$ ($\eta_{\mathrm{SD}}=0.182$). For this predefined cohort, attenuation therefore raises mean accuracy and substantially improves reproducibility, bringing the PLR-KAN result close to the power-law baseline ($0.7796\pm0.0084$).

High $I_p$ is complementary because ungated PLR-KAN already exceeds the power-law baseline ($0.6524\pm0.0712$ versus $0.6210\pm0.0050$). Gating further raises it to $0.6969\pm0.0426$ ($\Delta R^2=+0.0445\pm0.0576$). For PLR-MLP, the same operation raises $R^2$ from $0.4039\pm0.1417$ to $0.6510\pm0.1297$. These paired gains show that attenuation can help even when the KAN residual is already useful, while the branch-to-branch difference again emphasizes architecture dependence.

Smaller but consistent gains appear in the $P_L$ direction for the KAN branch ($\Delta R^2=+0.0086\pm0.0067$), while in the $\bar{n}_e$ direction the selection procedure itself recognizes that no protection is needed: the modal $\rho$ is $0$ for both residual branches (Table~\ref{tab:gate}), so many gated predictions are unchanged.

The stability effect is broader than the mean-score gains: $\eta_{\mathrm{SD}}<1$ in 11 of the 12 cohort--branch combinations in Table~\ref{tab:gate}. The post-hoc diagnostic nevertheless shows that statistical distance is not a universal monotonic proxy for residual error (Fig.~\ref{fig:distance_reliability}). For high $B_T$, the harmful-sample fraction rises from $73.9\pm13.6\%$ in the nearest quintile to $83.5\pm16.4\%$ in the farthest, but the mean within-pipeline Spearman coefficient is only $+0.137\pm0.260$, and neither binned error nor harmful fraction increases strictly monotonically. This is weak, direction-specific evidence rather than validation of a general distance--reliability law. For high $R$, the corresponding correlation is $+0.014\pm0.429$, and mean $q_i$ remains negative in all five quintiles. Statistical distance is therefore not equivalent to residual unreliability: samples can be far from the training distribution while still receiving useful residual corrections.

Reduced variability does not by itself guarantee higher accuracy, and high $R$ is the principal counterexample. Inner-edge calibration selects substantial attenuation ($\rho=0.50$), yet gated $R^2$ falls from $0.6384$ to $0.5986$ for PLR-KAN and from $0.6518$ to $0.5735$ for PLR-MLP because useful corrections are suppressed. On core-five, the KAN branch is marginally harmed ($0.9263\to0.9249$), whereas the MLP branch improves slightly ($0.9355\to0.9379$). Together with the high-$\epsilon$ failure, these counterexamples establish the scope of the third contribution: the gate is a task-calibrated empirical attenuation mechanism that can improve selected directions, not a universal out-of-distribution safeguard or a direct estimator of residual reliability. Physics- and regime-aware controls are required when statistical distance and residual usefulness diverge.

\subsection{Exploratory residual inspection}\label{subsec:pruning}

To inspect what the KAN residual branch can retain after sparsification, we use a separate exploratory interpolation run in which residual-amplitude regularization, internal KAN L1 and entropy penalties, and pruning are applied together, with empirical values $\lambda_r=0.05$ and $\lambda_1=\lambda_{\mathrm{entropy}}=1$. This setting differs from the main experiments, where $\lambda_r$ is selected within each pipeline and internal KAN sparsification is disabled. The visualization is therefore illustrative and is not used to support the comparative performance claims above.

Figure~\ref{fig:pruned} shows that the initially fully connected branch retains only a small number of effective connections, while several retained edge functions remain nonlinear. This representative run is consistent with the PLR design premise: after the power law captures the dominant trend, the residual branch can model structured, state-dependent deviations rather than only uncorrelated noise. The diagram does not establish that the same sparse topology or edge shapes recur across pipelines.

In this run, the $M_{\mathrm{eff}}$ input is removed under the chosen regularization. This is only a property of the displayed fit: it may reflect information already absorbed by the power-law anchor, correlations with other inputs, or the particular sparsification setting. It does not show that effective ion mass is physically irrelevant. Likewise, the learned edge shapes should not be extrapolated beyond the training support or interpreted as boundary-regime physics.

\onecolumngrid
\vspace{6pt}
\noindent\begin{minipage}{\textwidth}
    \centering
    \includegraphics[width=.50\linewidth]{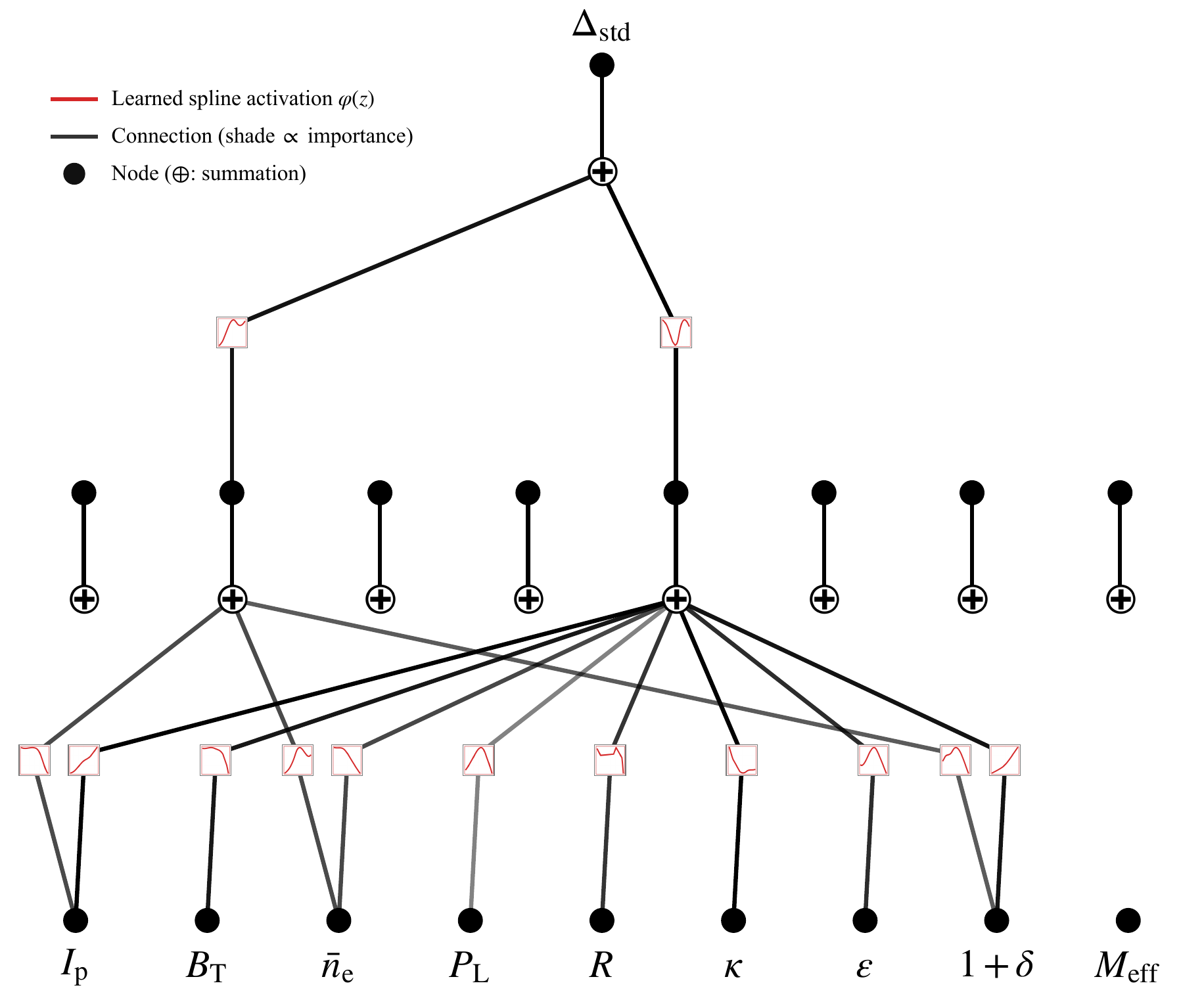}
    \captionof{figure}{Residual network structure of PLR-KAN after regularization and pruning in one representative exploratory interpolation run. The retained connections give a sparse nonlinear pathway from the inputs to the standardized residual $\Delta_{\mathrm{std}}$; red curves are learned edge activations $\varphi(z)$ and $\oplus$ denotes summation. The displayed topology is not an across-pipeline stability result.}
    \label{fig:pruned}
\end{minipage}
\par\vspace{6pt}
\twocolumngrid

\section{Conclusions}\label{sec:summary}

This work introduces a unified power-law-anchored residual-learning framework for H-mode energy-confinement prediction. A frozen empirical power law supplies the global trend, and a nonlinear learner is restricted to the systematic residual in logarithmic space. PLR-KAN is the primary high-capacity implementation, while the parameter-matched PLR-MLP replacement demonstrates that the formulation is not tied to KAN. On stratified random splits, both PLR variants nearly match the best direct predictor, showing that the anchor does not materially sacrifice interpolation accuracy.

The principal scientific result is that interpolation ranking does not determine behavior under parameter-defined shifts. Direct KAN gives the best interpolation score ($R^2=0.9680\pm0.0031$) but fails severely for high $B_T$ and is unstable in other directions. PLR-KAN retains comparable interpolation accuracy ($0.9671\pm0.0027$) and improves upon direct KAN in all five non-$\epsilon$ single-parameter-defined cohorts and core-five. PLR-MLP confirms the generality of the framework, while its direction-dependent comparison with PLR-KAN shows that anchoring effects remain architecture dependent. No formulation succeeds for $\epsilon>0.5$, where the held-out cohort is also a transfer to spherical-tokamak devices and a different confinement regime. Power-law anchoring therefore reduces severe unconstrained behavior but does not guarantee superiority to the power-law baseline or compensate for missing physics coverage.

The Mahalanobis prediction-time gate is an exploratory extension rather than the core contribution. It improves PLR-KAN for high $B_T$ from $R^2=0.4411\pm0.5029$ to $0.7595\pm0.0915$ and reduces its run-to-run standard deviation by approximately 82\%, but it degrades both residual branches for high $R$ and cannot repair the high-$\epsilon$ failure. These beneficial and harmful cases show that statistical distance can support task-calibrated residual attenuation in selected directions, but is neither equivalent to residual unreliability nor a universal out-of-distribution safeguard.

The present conclusions are bounded by the pooled DB5.2.3 engineering-variable dataset and by parameter-defined cohorts whose device compositions are correlated with their defining variables. Leave-one-device-out validation, external device datasets, dimensionless physics variables, and explicit predictive uncertainty are therefore important next tests.

The present scalar gate can be interpreted as a simple two-expert mixture between the empirical power-law baseline and the residual-corrected PLR prediction. A natural extension is a physics-aware mixture-of-experts framework~\cite{jacobs_adaptive_1991} in which multiple residual experts share a common power-law backbone but specialize in different device classes or operating regimes. A learned router could combine engineering variables, dimensionless physics parameters, device metadata, and predictive uncertainty to determine which residual correction is most trustworthy. Such an extension would require careful treatment of device imbalance, sparse regimes, and expert collapse.

\begin{acknowledgments}
The authors would like to thank the AI group of ENN Fusion for valuable discussions and helpful suggestions on this work.
This work was supported by the National Natural Science Foundation of China under Grant Nos. 12375215 and 12405275, and the Strategic Priority Research Program of the Chinese Academy of Sciences under Grant No. XDB0790201.

\end{acknowledgments}

\bibliographystyle{unsrt}
\bibliography{references}

\end{document}